\documentclass[prl,twocolumn]{revtex4}
\newcommand{\be}{\begin{equation}}
\newcommand{\ee}{\end{equation}}
\newcommand{\beqn}{\begin{eqnarray}}
\newcommand{\eeqn}{\end{eqnarray}}

\begin{document}

\title{{\bf LIV Dimensional Regularization and  Quantum Gravity effects  in
the Standard Model}}

\author{Jorge Alfaro}

\affiliation{Facultad de F\'{\i}sica, Pontificia Universidad Cat\'{o}lica de Chile \\
        Casilla 306, Santiago 22, Chile.
\\ {\tt jalfaro@puc.cl}}

\date{\today}

\begin{abstract}
Recently, we have remarked that the main effect of Quantum Gravity(QG)
will be to modify the measure of integration of loop integrals in a renormalizable Quantum Field Theory.  In the Standard Model this approach leads to definite predictions, depending on only one arbitrary parameter. In particular, we found that the maximal attainable velocity for particles is not the speed of light, but depends on the specific couplings of the particles within the Standard Model. Also
birrefringence occurs for charged leptons, but not for gauge bosons. Our predictions could be tested in the next generation of
neutrino detectors such as NUBE. In this paper, we elaborate more on this proposal.
In particular, we extend the dimensional regularization prescription to include Lorentz invariance violations(LIV) of the measure, preserving gauge invariance. Then we comment on the consistency of our proposal.
\end{abstract}

\maketitle


Recently, there has been a growing consensus that small LIV terms could  be the clue to obtain phenomenological predictions from Quantum Gravity \cite{1,2,3,4,5,6,7,8,9}. The reason for this is that a very tiny LIV produces a great effect on processes happening at energies much below the Planck scale.

In a previous letter\cite{alfaro} , we have remarked that the main effect of QG is to deform the measure of integration of Feynman graphs
at large four momenta by a tiny LIV. Within the Standard Model, such LIV implies several remarkable predictions, which are wholly determined up to an arbitrary parameter.Our predictions could be tested in the next generation of
neutrino detectors such as NUBE\cite{13,14}

{\bf LIV Dimensional Regularization}
We generalize dimensional regularization to a d dimensional space with an arbitrary constant metric $g_{\mu\nu}$. We work with a positive definite metric first  and then Wick rotate. We will illustrate the procedure with an example. Here $g=det(g_{\mu\nu})$ and $\Delta>0$.

\beqn
\frac{1}{g}\int\frac{d^dk}{(2\pi)^d}\frac{k_\mu k_\nu}{(k^2+\Delta)^n}=\nonumber\\
\frac{1}{g\Gamma(n)}\int_0^\infty dt t^{n-1}\int\frac{d^dk}{(2\pi)^d}k_\mu k_\nu e^{-t(g^{\alpha\beta}k_\alpha k_\beta+\Delta)}=\nonumber\\
\frac{g_{\mu\nu}}{2(4\pi)^{d/2}\Gamma(n)}\int_0^\infty dt t^{n-2-d/2} e^{-t\Delta}=\nonumber\\
\frac{1}{(4\pi)^{d/2}}\frac{g_{\mu\nu}}{2}\frac{\Gamma(n-1-d/2)}{\Gamma(n)}\frac{1}{\Delta^{n-1-d/2}}
\eeqn

In the same manner, we obtain, after Wick rotation:
\beqn
\frac{1}{g}\int\frac{d^dk}{(2\pi)^d}\frac{k_\mu k_\nu k_\rho k_\sigma}{(k^2-\Delta)^n}=
\frac{(-1)^n i}{(4\pi)^{d/2}}\frac{\Gamma(n-d/2-2)}{\Gamma(n)}\nonumber\\
\frac{1}{\Delta^{n-d/2-2}}
\frac{1}{4}(g_{\mu\nu}g_{\rho\sigma}+g_{\mu\rho}g_{\nu\sigma}+g_{\mu\rho}g_{\nu\sigma})\\
\frac{1}{g}\int\frac{d^dk}{(2\pi)^d}\frac{(k^2)^2}{(k^2-\Delta)^n}=
\frac{(-1)^n i}{(4\pi)^{d/2}}\frac{\Gamma(n-d/2-2)}{\Gamma(n)}
\frac{1}{\Delta^{n-d/2-2}}\nonumber\\
\frac{d(d+2)}{4}\\
\frac{1}{g}\int\frac{d^dk}{(2\pi)^d}\frac{k_\mu k_\nu}{(k^2-\Delta)^n}=
\frac{(-1)^{n-1} i}{(4\pi)^{d/2}}\frac{g_{\mu\nu}}{2}\frac{\Gamma(n-1-d/2)}{\Gamma(n)}\nonumber \\
\frac{1}{\Delta^{n-1-d/2}}\\
\frac{1}{g}\int\frac{d^dk}{(2\pi)^d}\frac{k^2}{(k^2-\Delta)^n}=
\frac{(-1)^{n-1} i}{(4\pi)^{d/2}}\frac{d}{2}\frac{\Gamma(n-d/2-1)}{\Gamma(n)}\nonumber \\
\frac{1}{\Delta^{n-d/2-1}}\\
\frac{1}{g}\int\frac{d^dk}{(2\pi)^d}\frac{1}{(k^2-\Delta)^n}=
\frac{(-1)^n i}{(4\pi)^{d/2}}\frac{\Gamma(n-d/2)}{\Gamma(n)}\nonumber \\
\frac{1}{\Delta^{n-d/2}}
\eeqn
We follow the notation of \cite{12}.

These definitions preserve gauge invariance, because the integration measure
is invariant under shifts. As a check, consider:
\be
K_\mu=\frac{1}{g}\int \frac{d^dk}{(2\pi)^d}[\frac{k_\mu+l_\mu}{[(k+l)^2-m^2+i0]^n}-\frac{k_\mu}{[k^2-m^2+i0]^n}]
\ee
to first order in $l_\mu$. With the definitions stated above, it is trivial to check that $K_\mu$ vanishes to first order in $l_\mu$.

To get a LIV measure, we assume that
\be
g_{\mu\nu}=\eta_{\mu\nu}+a\eta_{\mu 0}\eta_{\nu 0}\epsilon
\ee
where $\epsilon=2-\frac{d}{2}$. A formerly divergent integral will have a pole at $\epsilon=0$, so when we take the physical limit, $\epsilon->0$, the answer will contain a LIV term, precisely of the sort we obtain in \cite {alfaro}.

That is, LIV dimensional regularization consists in:

 1)Calculating the d-dimensional integrals using a general metric $g_{\mu\nu}$.

 2) Gamma matrix algebra is generalized to a general metric $g_{\mu\nu}$.

 3) At the end of the calculation, replace $g_{\mu\nu}=\eta_{\mu\nu}+a\eta_{\mu 0}\eta_{\nu 0}\epsilon$
 and then take the limit $\epsilon->0$.

As a concrete example, let us evaluate an integral that appears in the calculation of  the one loop results of \cite{alfaro}:
\beqn
A^{\mu\nu}=\int \frac{d^dk}{(2\pi)^d}\frac{k^\mu k^\nu}{[k^2-m^2+i0]^3}=\\
\frac{i}{(4\pi)^{d/2}}\frac{g^{\mu\nu}}{2}\frac{\Gamma(2-\frac{d}{2})}{2}\frac{1}{(m^2)^{2-\frac{d}{2}}}\\
=\frac{i}{(4\pi)^{d/2}}\frac{\eta^{\mu\nu}-a\delta^\mu_0\delta^\nu_0\epsilon}{2}\frac{\Gamma(2-\frac{d}{2})}{2}\frac{1}{(m^2)^{2-\frac{d}{2}}}\\
=\frac{i}{4(4\pi)^2}(\frac{\eta^{\mu\nu}}{\epsilon}-a\delta^\mu_0\delta^\nu_0) +{\rm a\  finite\  LI \ term}
\eeqn

From now on we take $\frac{a}{(4\pi)^2}=\alpha$.

Using LIV Dimensional Regularization we obtain the LIV photon self-energy in QED:
\beqn
L\Pi^{\mu\nu}(q)=
\frac{4}{3}e^2\alpha q_\alpha q_\beta\nonumber\\
(\eta^{\alpha\beta}\delta^\mu_0\delta^\nu_0+\eta^{\mu\nu}\delta^\alpha_0\delta^\beta_0-\eta^{\nu\beta}\delta^\mu_0\delta^\alpha_0-\eta^{\mu\alpha}\delta^\nu_0\delta^\beta_0)
\eeqn

LIV Dimensional Regularization reinforces our claim that these tiny LIV's originates in Quantum Gravity. In fact the sole change of the metric of space time is a correction of order $\epsilon$ and this is the source of the effects studied in \cite{alfaro}.
Quantum Gravity is the strongest candidate to produce such effects because the gravitational field is precisely the metric of space-time and tiny LIV modifications to the flat Minkowsky metric may be produced by quantum fluctuations.

{\bf Consistency of the approach}
In \cite{6} a Modified Standard Model with the LIV structure  of \cite{alfaro} has already been considered, albeit with arbitrary parameters. It was shown there that the theory is anomaly free and renormalizable: By rescaling coordinates and fields, the anomaly can be shown to be the same as in the Lorentz invariant situation, which is anomaly free. These results (anomaly free and renormalizability) also apply to our proposal, if we preserve gauge invariance, which is made easy by LIV Dimensional Regularization. In the  case\cite{alfaro}, at one loop the Standard Model predictions are not modified except for the LIV introduced in primitively logarithmically divergent integrals(which is determined uniquely up to a single constant, that measures the deformation of the space-time metric when we approach the physical dimension). Starting from the two loop contribution, in addition there will be also tiny LIV modifications to finite Standard Model predictions, as in \cite{6}.


\section*{Acknowledgements}
 The work of
JA is partially supported by Fondecyt 1010967 and Ecos-Conicyt C01E05. He wants to thank A.A. Andrianov for
several useful remarks. He wants to thank the hospitality of Ecole Normale Superieure,Paris. In particular
he wants to thank C. Kounnas, C. Bachas, V. Kazakov and A. Bilal for interesting discussions.



\begin{thebibliography}{99}

\bibitem{1} Amelino-Camelia, G.  et al., Tests of quantum gravity from observations of gamma-ray bursts,
Nature 393, 763 (1998).

\bibitem{2} Gambini, R.  and Pullin, J. , Nonstandard optics from quantum space time,
Phys. Rev. D 59, 124021 (1999).

\bibitem{3} Alfaro, J. Morales-T\'ecotl, H.A. and Urrutia, L.F. , Quantum gravity corrections to
neutrino propagation, Phys. Rev. Lett. 84, 2318 (2000).

\bibitem{4} Alfaro, J.  Morales-T\'ecotl, H.A.  and  Urrutia, L.F., Loop quantum gravity and light
propagation, Phys. Rev. D 65, 103509(2002).

\bibitem{5} Colladay, D.  and  Kostelecky, V.A.,  Lorentz-violating extension of the standard model,
Phys. Rev. D58, 116002(1998).

\bibitem{6} Coleman, S.and Glashow, S.L., High-energy tests of Lorentz invariance,
Phys. Rev. D 59, 116008 (1999).

\bibitem{7} Collins, J.  et al , Lorentz invariance and quantum gravity :an additional
fine-tuning problem?, Phys.Rev.Lett.93, 191301(2004).

\bibitem{8}T.Kifune, Astroph. J. Lett. {\bf 518}, 21 (1999);
G.Amelino-Camelia and T.Piran, Phys. Lett. B {\bf 497},  265
(2001);  J.Ellis, N.E.Mavromatos and D.V.Nanopoulos, Phys. Rev.
D {\bf 63}, 124025 (2001); G.Amelino-Camelia and T.Piran, Phys.
Rev. D {\bf 64}, 036005 (2001); G.Amelino-Camelia, Phys. Lett. B
{\bf 528}, 181 (2002).

\bibitem{9}J.Alfaro and G.Palma, Phys. Rev. D {\bf 65}, 103516 (2002);
J.Alfaro and G.Palma, Phys. Rev. D {\bf 67}, 083003 (2003).

\bibitem{alfaro} J. Alfaro, hep-th 0412295.

\bibitem{12}Peskin, M. and Schroeder D., An introduction to Quantum Field Theory,
Addison-Wesley Publishing Company, New York  1997.

\bibitem{13} Waxman,E. and Bahcall, J. ,High energy neutrinos from cosmological gamma
ray burst fireballs, Phys. Rev. Letts.78 , 2292(1997).

\bibitem{14} Roy,M. ,Crawford,H.J. and Trattner, A. , The prediction and detection of
uhe neutrino bursts, astro-ph/9903231.


\end{thebibliography}
\end{document}